\documentclass[amsmath,amssymb,twocolumn,amsfonts,twocolumn]{revtex4}
\usepackage{graphicx}
\usepackage{float}
\def \beq {\begin{equation}}
\def \eeq {\end{equation}}

\begin{document}
\title{ Comment on "Quantum Control and Entanglement in the Avian Compass"}
\author{I. K. Kominis}

\affiliation{Department of Physics, University of Crete, Heraklion
71103, Greece}

\maketitle
\noindent
{\bf (1)} It is well known \cite{geremia,boixo,romalis} that if one employs $n$ quantum systems to measure an energy splitting $E$ for time $T$, the precision is ($\hbar=1$) $\delta E=1/T\sqrt{n}$, where $\sqrt{n}$ is the "shot-noise-limited" signal-to-noise ratio of the measurement. If there are other noise sources, calling $(S/N)$ the signal-to-noise ratio, the measurement precision can be no better than $\delta E=1/T(S/N)$. The energy of electron spins in a magnetic field $B$ is $E=\gamma B$, where $\gamma=0.176~{\rm ns}^{-1}/{\rm mT}$, hence the smallest measurable field is $\delta B=1/\gamma T(S/N)$.\newline
{\bf (2)} Call ${\cal O}$ an observable which depends on the magnetic field. One can measure ${\cal O}$ to estimate $B$. If the measurement precision of ${\cal O}$ is $\delta{\cal O}$, then
\beq
\delta B={{\delta{\cal O}}\over {|\Delta{\cal O}/\Delta B |}}\label{sens}
\eeq
{\bf (3)} The authors in \cite{briegel} use a chemical reaction which lasts for a time $T_{r}$, the reaction time. The fundamental magnetic sensitivity limit is then \cite{note}
\beq
\delta B_{\rm fund}={{1/(S/N)}\over {\gamma T_{r}}}\label{msens}
\eeq
{\bf (4)} The authors introduce an observable ${\cal O}=T_{E}$, which is the entanglement lifetime, and the authors predict it's dependence on the
magnetic field, shown in Fig. 1a.
We can use this dependence to estimate $B$. To apply \eqref{sens} we need the precision $\delta T_{E}$ of measuring the entanglement lifetime $T_{E}$. Measuring time is like measuring frequency. If we measure a frequency $\nu$ during $T_{r}$, the  precision limit is, again, $\delta \nu=1/T_{r}(S/N)$. Setting $\nu=1/T_{E}$, it follows that $\delta T_{E}=T_{E}^{2}/(S/N)T_{r}$, leading to
\beq
\delta B_{T_E}={1\over {(S/N)}}{T_{E}^{2}\over T_{r}}{1\over {|\Delta T_{E}/\Delta B|}}\label{sensB}
\eeq
\newline
No matter how we perform the measurement, the sensitivity cannot surpass the fundamental limit \eqref{msens}. Thus the ratio $r\equiv\delta B_{T_E}/\delta B_{\rm fund}$ must be larger than unity, i.e.
\beq
r={{\gamma T_{E}^{2}}\over |\Delta T_{E}/\Delta B|}\geq 1
\eeq
\begin{figure}
\includegraphics[width=8.5 cm]{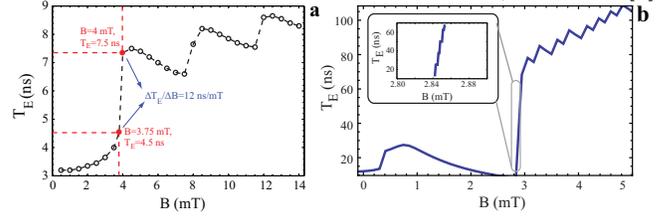}
\caption{(a) Fig. 2b of \cite{briegel}. (b) We do the same calculation for a radical-pair with one spin-1/2 nucleus
and an isotropic hyperfine coupling of 20 MHz. Zooming in at $B\approx 3$ mT it is evident that the slope $\Delta T_{E}/\Delta B$ becomes increasingly large, making $r<<1$, thus severely violating the fundamental limit \eqref{msens}.}
\end{figure}\noindent
{\bf (5)} For the hyperfine couplings used by the the authors the result for $r$ as follows from Fig. 1a is $r\approx 0.5$. However, if one zooms in at $B\approx 4$ mT where the discontinuity takes place, one finds that $r<<1$. In Fig. 1b we depict this using a very simple example of a radical-pair with just one spin-1/2 nucleus.
\newline
{\bf (6)} The root of the problem is the following.
The authors in \cite{briegel} made the fundamentally flawed assumption that the entanglement lifetime $T_E$ has nothing to do with the reaction time $T_r$. However, as shown by Kominis \cite{kominis_PRE} and Jones \& Hore \cite{JH}, the radical pair's spin coherence lifetime is on the order of $T_{r}$. The entanglement lifetime cannot be longer, as well demonstrated by the precision measurements community \cite{huelga}. Hence in reality $T_{E}$ is not a good magnetometric observable, since $T_{E}\approx T_{r}$ (i.e. the real $B$-dependence of $T_E$ is is wildly different from  Fig.1a) and $\Delta T_{E}/\Delta B\approx 0$, keeping $r\geq 1$.


\begin{thebibliography}{0}
\bibitem{note}
It is hopeless to improve on \eqref{msens} by waiting more than $T_r$, since the signal-to-noise ratio is exponentially
decaying with time, as the radical-ion pairs disappear with time constant $T_r$. This is easily proven formally by considering the maximum
of the function $f(t)=t\sqrt{n(t)}$, where $n(t)=n_{0}e^{-t/T_{r}}$ is the number of existing radical pairs. The maximum occurs at $t\approx T_{r}$, hence the minimum $\delta B$ is the one stated in \eqref{msens}.
\end{thebibliography}
\end{document}